# Stories and Systems: Educational Interactive Storytelling to Foster Media Literacy and Systemic Thinking

Christian Roth[1][0000-0001-7105-7758], Rahmin Bender-Salazar[2][0000-0002-5783-6314], Breanne Pitt[3][0000-0002-4987-5882]

[1] Professorship (Dis)connected Technology and Creativity, HKU University of the Arts Utrecht, Lange Viestraat 2b, Postbox 1520, 3500 BM Utrecht, Netherlands
christian.roth@hku.nl

[2] J.E. Cairnes School of Business & Economics, University of Galway, University Road, H91 TK33, Galway, Ireland
rahmin.bender@universityofgalway.ie

[3] School of Computer Science and Statistics, Trinity College Dublin, The University of Dublin, Dublin 2. D02 PN40 Ireland
pittb@tcd.ie

**Abstract.** In an era of super-complexity, where wicked problems resist straightforward solutions and the very frameworks for producing knowledge are contested, transformative learning is not a pedagogical option but a necessity. Without the capacity for systemic, transdisciplinary thinking, learners risk being overwhelmed by information complexity, susceptible to misinformation, and unable to act meaningfully on the challenges they face. This paper argues that Interactive Digital Narratives (IDNs) offer a distinctive pedagogical instrument for bridging this gap: by immersing learners in nonlinear, perspective-shifting narrative experiences, IDNs can cultivate the critical literacies and systemic sensibilities that transformative learning requires. However, realising this potential demands a principled framework for design and implementation. We introduce two connected contributions: the concept of Systemic Learning IDNs, which are narrative experiences explicitly designed to support learners in exploring complex systems, and the CLASS transdisciplinary pedagogical framework, a structured model integrating systems thinking, design thinking, and interactive storytelling to guide their creation and use. Applying CLASS to two contrasting case studies, a commercial narrative simulation and a student-developed educational prototype, we demonstrate how IDNs can be designed and deployed to foster media literacy, systemic thinking, and collaborative problem-solving across disciplinary contexts. The paper offers both conceptual grounding and practical recommendations for educators seeking to implement transformative, systems-oriented learning in an increasingly complex world.

**Highlighted contributions:**

- We examine the diverse theoretical foundations underpinning a transdisciplinary approach to integrating media literacy, IDNs, and systems thinking in educational settings.
- We introduce the concept of *Systemic Learning IDNs:* Interactive Digital Narrative experiences explicitly designed to support learners in exploring, understanding, and reflecting on complex systems and interdependencies.
- We propose the CLASS transdisciplinary pedagogical framework, a structured model intended to guide the design, implementation, and evaluation of Systemic Learning IDNs. While the framework has relevance for a broad range of stakeholders, including designers, educators, and researchers, this paper focuses on its application in classroom settings and demonstrates its transferability across disciplinary contexts.
- We apply CLASS to two contrasting case studies, *Suzerain* and *Bezmarisk*, to demonstrate the framework's utility across a spectrum from a commercially available narrative simulation to a purpose-built educational prototype. We analyse and compare both through the CLASS framework and offer practical recommendations for classroom integration across different disciplinary settings.

## 1. Introduction

The internet has granted learners unparalleled access to diverse narratives and an abundance of information. Yet, many of today's most pressing challenges, such as climate change, global pandemics, and social inequality, are wicked problems (Churchman, 1967; Rittel & Webber, 1973; Bender-Salazar, 2023; Grewatsch et al., 2021): deeply entangled, cross-cutting issues that resist straightforward solutions. These problems traverse ecological, social, and economic systems, often overwhelming efforts to locate trustworthy information or reach consensus on effective responses.

While digital technologies and global connectivity have increased access to knowledge, they have also exacerbated fragmentation. Learners are confronted with an overwhelming volume of competing narratives, making it increasingly difficult to assess credibility, contextualise insights, and translate them into action. Simultaneously, mass media and algorithmic social platforms tend to privilege engagement over accuracy, amplifying misinformation, disinformation, and confirmation bias. The resulting public discourse around complex issues is often simplified, polarised, or incoherent.

Barnett (2000) refers to this evolving educational landscape as one of super-complexity: a condition marked not only by uncertainty and unpredictability in the external world, but also by the contestability of the frameworks through which knowledge is produced and legitimised. Wicked problems—ranging from technological disruption to global inequality—exemplify this super-complexity and call for educational approaches that transcend reductive content delivery. While prior work has explored IDNs, media literacy, and systems thinking separately, there is limited work that integrates these into a coherent pedagogical framework for transdisciplinary learning. This paper addresses that gap with the CLASS framework by integrating systems thinking, design thinking, and interactive storytelling. It supports learners in cultivating the cognitive, ethical, and collaborative competencies needed to navigate a fragmented and volatile world. In doing so, it aligns with Barnett's vision of epistemic resilience, preparing learners to act meaningfully under conditions of uncertainty.

Transdisciplinary learning does not replace discipline-specific knowledge but builds on and extends it by creating connections across domains with theory and practice. The CLASS framework reflects this: it synthesises established learning theories, including constructivism (Piaget, 1952; von Foerster, 2003), experiential learning (Kolb, 1984), scaffolded knowledge-building (Vygotsky, 1978), and double-loop reflection (Argyris & Schön, 1978), with the specific affordances of interactive narrative design. Central to this synthesis is the role of curiosity and creativity as drivers of engagement and deeper learning. Rather than positioning transdisciplinary approaches as opposed to disciplinary thinking, CLASS treats the interplay between them as generative, helping learners move between focused, discipline-specific inquiry and broader systemic understanding. At the same time, this paper acknowledges the real tension: dominant educational models remain largely organised around discrete subject areas, and introducing transdisciplinary approaches requires deliberate design within those existing structures. While transdisciplinary approaches offer powerful opportunities for integrating systems thinking, storytelling, and media literacy, they also face significant challenges within existing educational structures. These include the risk of superficial integration across disciplines, institutional resistance due to rigid curricula, and difficulties in assessing complex, process-oriented learning outcomes. As such, the implementation of frameworks like CLASS requires careful alignment with disciplinary goals, assessment practices, and educator support to ensure meaningful and sustainable impact.

Interactive Digital Narratives (IDNs) (Koenitz et al., 2022; Koenitz, 2023; Roth, 2019; Roth, Pitt, Bender-Salazar, 2024) offer a promising pedagogical instrument in this context. IDNs allow learners to explore complex issues through multiple roles and perspectives within nonlinear narratives. Central to this is the notion of virtual identity, defined by The Philosophy of Technology and Engineering Sciences (2009) as social identities assumed or presented by individuals in computer-mediated environments. IDNs offer structured yet flexible spaces for learners to construct, negotiate, and reflect on such identities, facilitating exploration not only of external roles, but also of the digital self.

Janet Murray's (1997) concept of Scripting the Interactor (StI) exemplifies how interactive storytelling can invite active participation by contextualising a role, managing expectations, and surfacing actionable choices. This approach supports a constructivist paradigm in which learners engage more deeply with content through role identification and experiential meaning-making (Pitt & Roth, 2023).

Murray's framework further identifies four key affordances of digital media: procedural (rule-based systems), participatory (inviting user interaction), encyclopaedic (offering vast multimedia information), and spatial (creating navigable virtual environments); alongside three experiential qualities: agency (the ability to take meaningful action), immersion (a sense of transportation into a simulated space), and transformation (the integration of enacted experiences into personal meaning). These qualities offer a high-level structure for understanding learner experience in IDNs.

However, to fully realise the transformative potential of IDNs, learners must develop the critical skills to engage with and evaluate the content they encounter. This calls for a shift toward new literacies (Sang, 2017; Shahzadi & Surif, 2023), a cluster of digital, media, information, internet, and computer literacies (Coiro et al., 2014), that collectively shape how individuals interact with and interpret information in a digital world.

Among these, media literacy plays a central role. In a media-saturated culture, media functions as a form of public pedagogy, shaping how information, values, and ideologies are disseminated and internalised (Share, Mamikonyan, & Lopez, 2007). Developing media literacy enables learners to critically analyse, assess, and create media with awareness and purpose.

Yet, literacy alone is insufficient for grappling with wicked problems. Learners must also develop systems thinking—the ability to perceive patterns, relationships, and feedback loops within complex systems (Dentoni et al., 2023; Meadows, 2008; Senge, 1996; Stroh, 2015). Practices such as systems mapping (Senge et al., 2015) help learners visualise interconnections, anticipate consequences, and adopt multiple perspectives.

Bringing together Systems Thinking and Mapping (Dentoni et al., 2022; Roth, Pitt, Bender-Salazar, 2024) with Interactive Digital Narratives creates a transdisciplinary pedagogical framework for collective sensemaking. IDNs, when grounded in real-world complexity, enable learners to explore, visualise, and engage with the interdependencies that characterise wicked problems. This integrated approach fosters both cognitive and affective capacities, encouraging learners to reflect, collaborate, and act with greater intentionality.

This conceptual paper makes three connected contributions. First, it synthesises the theoretical foundations of a transdisciplinary approach to IDN-based education, drawing on systems thinking, design thinking, and interactive storytelling. Second, it proposes the CLASS transdisciplinary pedagogical framework as a structured guide for the design, implementation, and evaluation of Systemic Learning IDNs. Third, it applies CLASS analytically to two contrasting case studies, a commercially available narrative simulation and a purpose-built educational prototype, to demonstrate the framework's utility across a spectrum of complexity, accessibility, and disciplinary context, and to derive practical recommendations for classroom integration.

A note on terminology: this paper uses several related but distinct pedagogical terms, and it is worth clarifying their meaning as used here. *Pedagogy* refers broadly to the theory and practice of teaching and learning. The CLASS framework is a *pedagogical framework*, a structured, principled model that guides the design, implementation, and evaluation of learning experiences. It is informed by a *pedagogical approach*: a broad orientation drawing on transdisciplinary, experiential, and constructivist traditions. Within the framework, the five levers (curiosity, lens & scope, agency, scaffolds, sandboxes) function as *pedagogical principles*, the underlying commitments that shape learning design. These are enacted through *pedagogical strategies* (e.g. pre-IDN mapping exercises, role-based debriefs), which in turn draw on specific *pedagogical methods* (e.g. causal loop mapping, reflective journaling) and *pedagogical instruments* (e.g. policy cards in *Bezmarisk*). The IDN itself functions as the *pedagogical medium*, the converging form through which tools, methods, and strategies operate, rather than a discrete instrument in its own right. Where this paper refers to broader institutional or curriculum-level structures, the term *educational framework* is used, distinct from CLASS itself.

## 2. Core components of the pedagogical approach

This section outlines the foundational components that underpin the proposed pedagogical approach: systems thinking, design thinking, and storytelling. Each allowing for a distinct yet complementary approach to understanding and navigating complexity. Systems thinking enables learners to identify interconnections and feedback loops in dynamic systems; design thinking fosters iterative, empathy-driven problem solving; and storytelling engages learners emotionally and cognitively through narrative exploration. Together, these components form the theoretical and practical basis for integrating IDNs into transformative learning experiences for complex topics.

### *2.1 Systems thinking: Understanding complexity through interconnections*

Systems thinking (Dentoni et al., 2023; Meadows, 2008; Senge, 1996; Stroh, 2015) is a framework for understanding dynamic relationships between interdependent elements in social and ecological systems—and a crucial skill for navigating the grand challenges these systems face. Building on general systems theory (von Bertalanffy, 1968), it defines systems as foundational models of organisation between parts that form a cohesive and relational whole, in which changes in one part create ripple effects across the entire system (Meadows, 2008). In a world of entrenched and interdependent social, ecological, economic, and justice systems, systems thinking provides the ability to see these interconnections, anticipate consequences, and navigate complex terrain (Senge, 1990).

Systems mapping is a method to visualise and co-create an artefact of the mapping process and the system as imagined in that specific setting (Dentoni et al., 2022). Embedding systems thinking and mapping into formal and informal education could allow more learners to integrate these skills into their team processes. Systems thinking and mapping education can also be useful for individuals and organisations gaining understanding and addressing global challenges (Bender-Salazar & Dentoni, 2020). By teaching students to see the interconnectedness of social, environmental, and economic systems, we empower them to think holistically and act effectively. Real-world problems are rarely simple, straightforward and linear in nature. Wicked problems require multivariate analysis, which considers multiple variables and their interactions, and enables a more accurate understanding of complex issues. Systems thinking can help students grasp the intricate relationships between various factors, stimulating curiosity and enhancing their ability to tackle complex problems.

### *2.2 Design thinking and designerly learning*

In parallel with systems thinking and mapping, design thinking offers a complementary method rooted in creativity, iteration, and reflective learning. A participatory design approach (Spinuzzi, 2005; Cumbo & Selwyn, 2021) brings together educators, students, and practitioners to collaboratively design learning environments that address transdisciplinary and real-world challenges. These design methodologies ensure that educational content is not only relevant and meaningful but also fosters greater engagement, motivation, and a sense of ownership among learners.

In this research, we draw on design thinking as a form of “designerly ways of knowing and doing” (Self et al., 2013; Cross, 2011) a reflective, iterative, and human-centred approach increasingly used to address complex social and ecological issues (Bender-Salazar, 2023). When combined with systems thinking, design thinking enables learners to both zoom in to understand the details of a problem and zoom out to grasp its broader systemic context (Dentoni et al., 2022). This dual capacity supports the iterative development of potential solutions while maintaining awareness of interdependencies and long-term impacts.

### *2.3 The power of interactive stories in education*

Storytelling has historically been a vital tool for understanding and sharing knowledge, and researchers such as Roger Schank (1990; 1999) and Jerome Bruner (1990; 1991) theorised its central importance to human learning, thinking, and understanding. Stories offer an impactful instrument for teaching skills to solve challenging problems (Jonassen & Hernández-Serrano, 2002). IDNs (Koenitz, 2023; Roth, 2019) extend this tradition into interactive, immersive forms, fostering self-directed learning by allowing learners to make meaningful choices and explore complex ideas (Murray, 1997; Ryan, 2006; Roth & Koenitz, 2016). Educators can benefit from this interactive technology to engage students and enhance their understanding of complex social and ecological issues (Roth et al., 2023).

However, digital technology alone cannot spark curiosity, learning, and creativity. Thoughtful stories based on real-world problems, showing systems interactions, challenges, and socio-ecological issues, must serve as scaffolding for learners, guiding their understanding and helping them make sense of the complexity that digital technologies aim to represent (Pitt & Roth, 2023). Commercial interactive narrative experiences meant for entertainment already illustrate how IDNs can simulate complex real-world challenges. *This War of Mine* places players in the midst of moral dilemmas faced by civilians in war-torn environments, while *Frostpunk* challenges them to navigate the tension between survival and ethics in a frozen dystopia. However, despite their pedagogical potential, both titles are often too time-consuming and emotionally intense for practical use in classroom settings. To support education directly, we need specially created, balanced systemic learning IDNs that can serve as experiential learning interventions. These condensed experiences can encourage critical thinking, spark meaningful conversations, and be tailored to specific learning objectives, fitting within class periods or homework schedules.

## 3. Integrating framework and practice

To create a transdisciplinary pedagogical framework, we integrate systems thinking, design thinking, and interactive storytelling.

### *3.1 Systems thinking and design thinking as interlocking tools*

Systems thinking helps uncover the intricate relationships and feedback loops that shape wicked problems. Systems thinking, in the context of learning, can be considered a skill and ability that allows a problem's full setting, interconnecting elements, and frameworks to be seen in order to understand better dynamics and relationships (Hosley et al., 1994). Systems thinking can thus be an impactful tool for organisational learning and problem solving (Fullan, 2005; Senge, 2000) when integrated into an artefact that serves as a boundary object (Black, 2013; Nathues et al., 2024) between users to have shared understanding of the elements and relationships involved. This ability for systems thinking to help the learning and understanding of users can be particularly useful in learning and also in integrating concepts of design thinking. Design thinking provides a framework for experimentation, empathy, and innovation for approaching wicked problems. Design thinking for learning and addressing wicked problems can aid learners to delve into complexity and gain a deeper understanding (Bender-Salazar, 2023), if the approach is not oriented towards solution finding and crafting solvable problems (Richterich, 2023). When combined in educational practice, systems thinking and design thinking operate as interlocking lenses. Systems thinking provides the analytical scaffolding, helping learners map interdependencies, identify feedback loops, and situate individual decisions within broader structures. Design thinking then activates that understanding through action: it moves learners from analysis into iterative experimentation, empathy-driven problem framing, and the prototyping of potential responses. Together, they support a form of learning that is simultaneously expansive and practical, capable of holding complexity while still generating meaningful next steps. This integration is particularly productive when embedded in a shared artefact or boundary object (Black, 2013; Nathues et al., 2024) that allows diverse stakeholders to co-construct understanding of a system's elements and relationships. Crucially, design thinking in this context must resist the pull toward solutionism (Richterich, 2023), the tendency to reduce wicked problems to solvable puzzles, and instead remain open to ambiguity and ongoing reframing.

### *3.2 Storytelling as humanising systemic learning*

Storytelling serves as the connective tissue between systems and design thinking as it humanises the learning experience, giving form to abstract systems and allowing learners to explore the consequences of different choices. When learners not only read or observe but also *perform* within a narrative, through role-play, collaboration, and interaction with other participants, they actively co-construct meaning. By stepping into diverse roles and relating to characters and situations, they are encouraged to explore perspectives that differ from their own. This embodied and social engagement fosters curiosity, emotional resonance, and potentially empathy, resulting in complex responses when engaging with human-centred, systemic issues.

However, while empathy can serve as a gateway to understanding others' experiences, it also requires critical examination (Rouse, 2021). In complex, interconnected scenarios, premature or unexamined empathy can reinforce biases or oversimplify conflicting perspectives. Therefore, storytelling in educational IDNs must not only evoke empathy but also support learners in *critically reflecting* on their emotional responses, helping them recognise when empathic alignment might obscure systemic dynamics or power imbalances.

In this way, storytelling, when combined with systems and design thinking, invites learners to navigate between emotional engagement and analytical distance. This balance equips them to engage more responsibly and creatively with global challenges while remaining aware of the limitations of their own perspectives.

### *3.3 Interactive Digital Narratives (IDNs) as a converging medium*

If these elements are then integrated into an interactive digital story—enhanced with visual, audio, and interactive elements—they can be a tool that can be used across learning environments around the world. IDNs can provide a unique platform for systems thinking and mapping, design thinking, and storytelling to combine forces, allowing learners to visualise and navigate topics and issues in ways that traditional education methods cannot alone. Recent work has demonstrated this potential empirically: Pitt, Youngblood, and Haahr (2026) show how an educational IDN grounded in systems thinking can foster ideological flexibility and perspective-shifting in secondary learners, using a non-playable character as a More Knowledgeable Other (MKO) to scaffold engagement with complex viewpoints. This combination of transdisciplinary skills and abilities is fostered by a Hybrid Design Thinking Model (Bender-Salazar, 2023) that integrates learning sciences, design methodologies and systems orientation.

### *3.4 Experiential and reflective learning integration*

The integration of transdisciplinary approaches in IDNs aims to help learners better understand and engage with wicked problems This approach helps learners move between the concrete and the conceptual elements of a problem while experimenting with ideas and reflecting on this process, which can be described as Kolb's experiential learning model (1984). This model emphasises the need for learners to include concrete action, learning from experience, reflection, and experimentation and flow between them as they learn (Bender-Salazar, 2023; Beckman & Barry, 2007). Along with the Kolb experiential learning model, this transdisciplinary approach to IDN education fosters *double-loop learning* (Argyris & Schön, 1978), defined as a process of approaching problems that emphasises learning from mistakes and reflection on approaches to refine future actions. This iterative process of reflection and revision can also be found in the concept of the *hermeneutic circle* (Gadamer, 2004), the idea that understanding a part of a text or experience depends on understanding the whole, and vice versa. Meaning emerges through an ongoing back-and-forth between parts and the whole. In narrative interpretation, this means that earlier events are re-evaluated in light of new information, leading to a continual reinterpretation of the entire experience.

In the context of interactive digital narratives (IDNs), this dynamic is further complicated by interactivity and player agency, which introduce a second interpretative layer. This has been described as a *double-hermeneutic circle* (Karhulahti, 2012; Roth et al., 2018) or *hermeneutic spiral* (Knoller, 2019), where the user's choices not only influence the unfolding narrative but also reshape their understanding of both prior events and the system itself. Such engagement fosters a form of experiential learning that is both reflective and emergent, reinforcing the conditions for double-loop learning.

In this way, IDN education not only supports technical or artistic competence but also cultivates critical awareness, adaptive thinking, and the capacity to question underlying assumptions; skills essential for navigating complex, real-world challenges. This combination of experiential and reflective learning strategies forms the structural foundation for IDN-based education designed to cultivate systems and design thinking skills.

While interactive digital narratives (IDNs) offer learners the opportunity to engage with complex systems and media structures through exploration and agency, it is essential to avoid conflating interactivity with critical awareness. As Žižek (1999) argues, the illusion of choice and agency can often reinforce the very ideological structures users believe they are navigating or questioning. Within media and system literacy education, this insight serves as a crucial reminder that the structure of an IDN is never neutral: the logic of available choices, narrative branches, and constraints reflects and often obscures underlying assumptions and value systems. Similarly, ethical considerations must be integrated throughout both the design and implementation phases (Koenitz, Barbara, & Bakk, 2022). The role of the educator, then, becomes central: not only to support comprehension and interpretation, but to help learners interrogate the affordances, blind spots, and ideological implications of the interactive structure itself.

### *3.5 Building knowledge through scaffolded exploration*

Extending the foundation of experiential and reflective learning, Interactive Digital Narratives (IDNs) provide a sandbox for active knowledge construction. Their design inherently aligns with leading educational theories such as constructivism and Vygotsky's Zone of Proximal Development (ZPD), offering learners scaffolded, guided experiences that support deep engagement with complexity.

The learning theory of constructivism (Piaget, 1952) suggests that learners actively construct meaning based on their interactions and interpretations, rather than passively receiving objective knowledge. IDNs exemplify this principle by placing learners in narrative-driven environments that promote agency, reflection, and contextual sense-making. Constructivist learning, particularly as influenced by second-order cybernetics and radical constructivism, further recognises that knowledge is situated, shaped by the observer's context, and continuously developed through experience (von Foerster, 2003; Glanville, 2010). However, effective constructivist learning requires appropriate levels of guidance, and IDNs must be intentionally designed to balance autonomy with structured support.

This is where Vygotsky's (1978) Zone of Proximal Development (ZPD) becomes relevant. Learners can achieve higher cognitive levels with support from a MKO (Figure 1). In systemic learning IDNs, the MKO may be embedded in the narrative design—via characters, tasks, or feedback—but equally importantly, this role is often fulfilled by the facilitator or teacher. This embedding of the MKO within narrative design has been demonstrated empirically by Pitt,

Youngblood, and Haahr (2026), whose educational IDN uses a non-playable character as MKO to guide learners through ideologically complex scenarios, with MKO engagement found to be a strong predictor of perspective change. The facilitator serves not only as a facilitator of the IDN experience but also as a critical guide who can adapt scaffolding to student needs, clarify points of confusion, and provide emotional or conceptual support when learners encounter challenges beyond their immediate grasp. The facilitator acts as both co-navigator and sounding board, adapting scaffolding to student needs and helping learners make sense of ambiguity and complexity through dialogue and reflection.

This framing also highlights that while IDNs can be a potent ingredient in the "learning meal," they are not a standalone solution. Deep learning often emerges from the interaction between learner and teacher, through questioning, contextual framing, and pre- and post-IDN discussion. The collaborative dynamic between learners and educators transforms narrative exploration into sustained inquiry, where interpretation and insight are co-created and reinforced.

Together, these design and learning principles position IDNs as both constructivist tools and scaffolding mechanisms that support learners in building, testing, and refining mental models of complex systems. Through such scaffolded exploration, enriched by the presence of a responsive facilitator, learners can internalize lessons from the narrative experience and transfer them to real-world contexts, reinforcing both cognitive and affective dimensions of systems and media literacy.

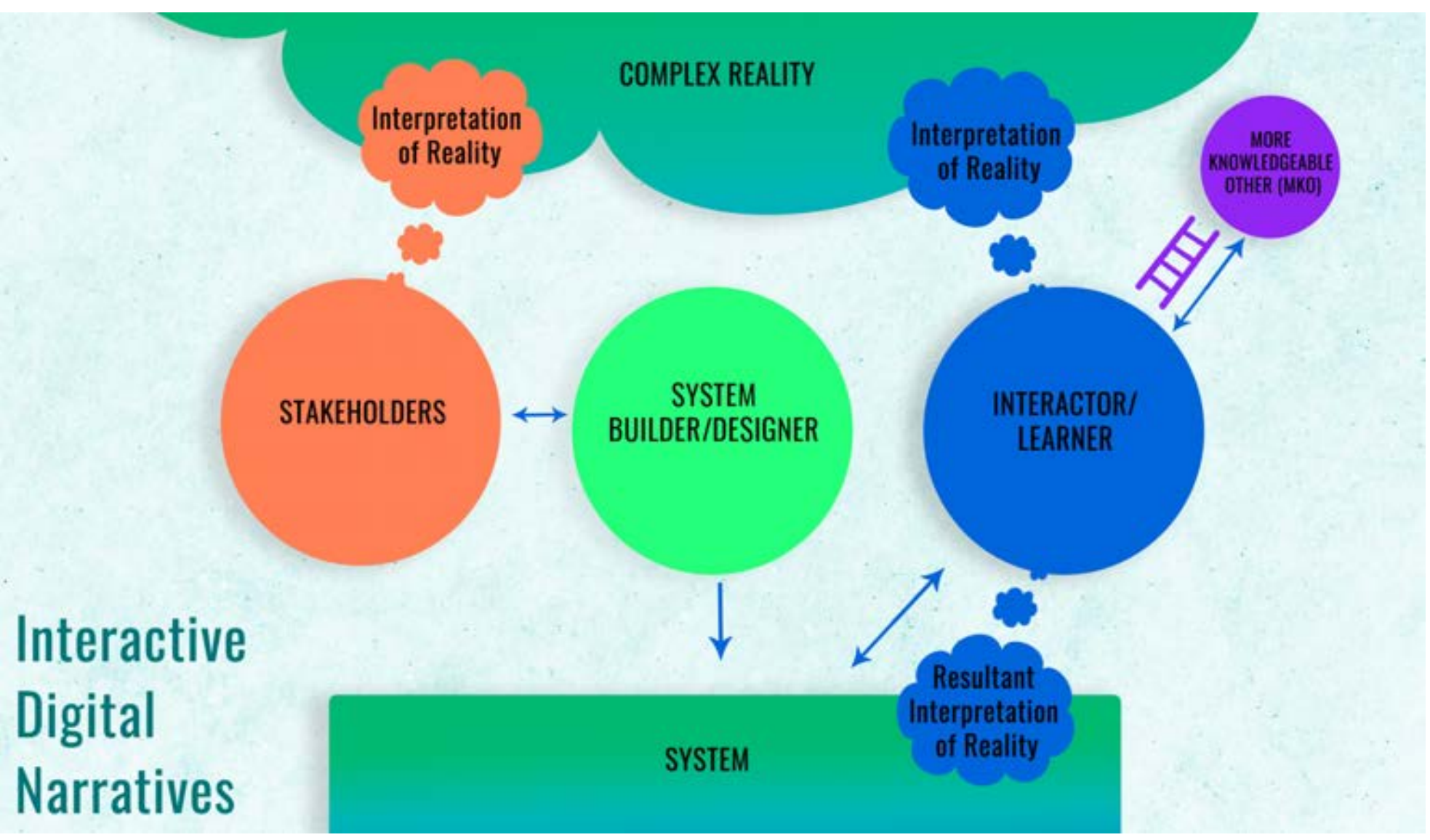

Figure 1: Visual representation of how a learner's interpretation of reality is shaped through interaction with the IDN system, supported by a More Knowledgeable Other (MKO) when necessary to guide understanding and scaffold meaning-making.

### *3.6 Systemic Learning IDNs as a response to super-complexity*

Barnett (2000) argues that traditional models of education, which aim to transmit fixed bodies of knowledge, are increasingly inadequate in a world marked by *super-complexity*, a condition in which not only are problems ill-defined and ever-changing, but the very frameworks for interpreting and responding to them are themselves contested. In contrast to complexity, which assumes that challenges may still be resolved through sufficient information and expertise, super-complexity calls into question the legitimacy, coherence, and stability of knowledge itself. In this context, higher education must move beyond content delivery and instead cultivate in learners the capacity to live purposefully, ethically, and critically amidst uncertainty. Barnett calls for an education that equips learners not only with what is *known*, but with the dispositions and capabilities to question, navigate, and co-create meaning in ambiguous and shifting environments. This entails fostering epistemic humility, critical reflexivity, openness to plural perspectives, and imaginative engagement with real-world dilemmas.

Systemic learning IDNs, as proposed in this paper, offer an impactful pedagogical response to this call. By embedding learners within dynamic, interactive stories that simulate wicked problems—such as misinformation, climate breakdown, and algorithmic governance—these experiences create scaffolded environments in which learners are invited to experiment with multiple perspectives, engage ethically with consequences, and reflect critically on systems of power, knowledge, and identity. In doing so, systemic learning IDNs support not only cognitive learning, but the kind of ontological development Barnett advocates: helping students become more resilient, responsible, and reflective participants in an increasingly contested world.

### *3.7 Practical considerations and classroom integration*

While theoretically robust, implementing IDNs in real classrooms presents logistical challenges. Educators often face rigid curricula and limited time for supplemental materials. To address this, IDN interventions must be adaptable and easy to integrate into existing educational frameworks. In practice, this means designing scaffolded experiences that visualise complex systems and ethical tensions, supported by collaborative and reflective activities. Preparing learners before the IDN intervention, through pre-reading on systemic topics, role assignments, and introductory mapping exercises, activates prior knowledge and sets the context. After playing, activities such as causal loop mapping, role-based debriefs, and reflective discussion support deeper processing of the narrative experience. For educators, effective implementation requires time-efficient materials, clearly defined learning outcomes, modular content, and facilitation guides that allow the framework to flex around existing objectives.

## 4. The CLASS Transdisciplinary Pedagogical Framework: Five levers of learning

The CLASS framework integrates systems thinking and mapping, design thinking, and storytelling with IDNs to create an engaging and holistic learning environment. By leveraging storytelling, educators can inspire curiosity and creativity, preparing learners to tackle global challenges in a holistic way.

This framework is built around five key components that foster a comprehensive educational experience (Bender-Salazar et al., 2024). It describes a structured model to guide the design and evaluation of IDN-based learning interventions, and consists of five interconnected educational levers as shown in Figure 2. The first lever is creative curiosity, in which learners can explore their interests, ideas, challenges, and questions of their world using arts, imagination, and creative ideation. This lever draws on Loewenstein's (1994) information gap theory of curiosity, which holds that curiosity is triggered when learners become aware of a gap between what they know and what they want to know. IDNs are particularly well suited to activating this drive: branching narratives, withheld information, and meaningful choices create precisely the conditions of uncertainty and anticipation that motivate learners to explore further. The second lever of learning is lens and scope, which ensures learners are exposed to diverse thoughts, perspectives, historical context, and interconnections to deepen their understanding of complex issues and problems. The third lever is agency , which enables the learners to make significant choices and direct their educational pathway, cultivating empowerment, influence, and responsibility. The fourth lever is scaffolds, which refers to the support and guidance customised and tailored to each student to aid but not interfere with their learning journey. Finally, the fifth lever is sandboxes, referring to the physical and virtual learning space in which the learner operates, where they can safely take risks and play with ideas and concepts. These five levers, as shown in Figure 2, are the responsibility of the educators and the learner and are applied in no particular order or importance, as they all work in concert.

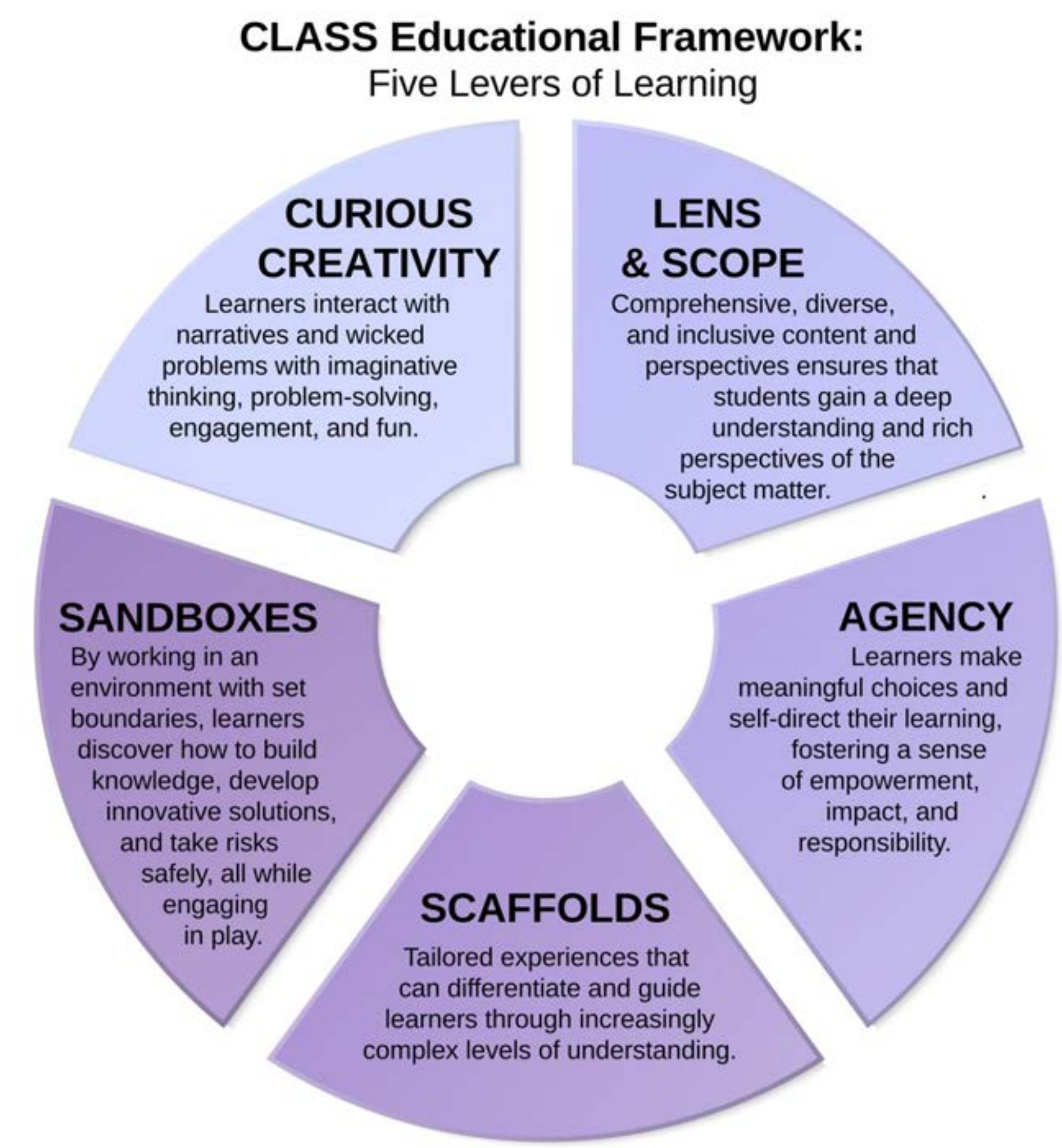

Figure 2: CLASS Framework with 5 levers of learning

This framework is meant to guide the design and analysis of systemic learning IDNs by bringing together different stakeholders, such as educators, developers, learners, and subject matter experts, fostering a collaborative environment where each group contributes their expertise. Educators can provide insights into pedagogical goals and learning outcomes, while developers can create the interactive and technical elements that make the narratives engaging (cf. Roth, Pitt, Bender-Salazar, 2024). As users, learners can offer valuable feedback to ensure the narratives are engaging, accessible, and aligned with their learning needs. Subject matter experts can help ensure the accuracy and relevance of the content, grounding the stories in real-world contexts.

While the case studies in this paper focus on civic education and media literacy, the five CLASS levers are discipline-agnostic: they describe conditions for learning rather than content. This means the framework can guide the design and implementation of Systemic Learning IDNs across a wide range of subject areas. In environmental science, for example, IDNs can simulate ecological decision-making and the long-term consequences of resource management, drawing directly on the systems mapping approaches already embedded in CLASS. In health education, role-based narrative simulations can place learners in ethical dilemmas around patient care or public health policy, activating the agency and sandbox levers in contexts where real-world consequences are high. In history, the "what if" structure of branching narratives supports counterfactual inquiry and perspective-taking across different historical actors and periods (Baldwin & Ching, 2017). In business and organisational learning, IDNs can model leadership decisions and their systemic ripple effects, building on the double-loop learning and systems thinking traditions already present in the CLASS theoretical foundation (Argyris & Schön, 1978; Senge, 1990). Across all these contexts, what transfers is not the content of the case studies presented here, but the CLASS framework itself as a structured guide for selecting, designing, and evaluating IDN-based learning experiences.

## 5. Illustrative Case Studies

To illustrate the practical application of the integrated pedagogical approach and the CLASS framework, this section presents two case studies involving IDNs. These examples demonstrate how systems thinking, design thinking, and storytelling can be operationalised within educational contexts to promote media literacy. The first case explores the commercially developed game *Suzerain*, while the second examines *Bezmarisk*, a student-created role-based learning activity. Collectively these case studies offer insight into how IDNs can scaffold complex learning experiences across diverse classroom settings.

### *5.1 Case Study 1*

We sought an existing playful narrative simulation to map onto our CLASS framework. Roth, Knoller, Haak, and Lund (2022) propose political interactive narrative simulations as a means to introduce complexity into the classroom context. Adopting this approach, we identified a relevant experience for analysis: *Suzerain* (Torpor Games, 2020) provides a strong match for all five dimensions, especially when used in educational contexts that aim to build media literacy, ethical reasoning, and systemic insight.

#### *5.1.1 Introduction: Suzerain as Media Literacy and Systemic Thinking training*

Suzerain is a narrative-driven political simulation game that places players in the role of President Anton Rayne, the newly elected leader of the fictional Republic of Sordland. Set in a post-civil war context facing democratic reform, the game engages players in morally complex decision-making that affects the nation's political, economic, and social future. Through branching narratives and systemic dilemmas, Suzerain exemplifies the CLASS transdisciplinary pedagogical framework, integrating IDNs and systems thinking and thus creating a potentially impactful learning environment.

This aligns closely with Peter Senge's (1990) concept of systems thinking as the ability to see the whole, identify causal relationships, and recognise delayed or distant consequences. As players revise their strategies in response to emergent outcomes, they move beyond linear cause-and-effect thinking into a more dynamic understanding of governance, policy, and communication.

Unlike traditional media education, which often teaches learners to deconstruct texts from the outside, *Suzerain* requires them to engage *from within* the system. The player must interpret conflicting news sources, deal with propaganda, and decide whether to support or suppress freedom of the press. The media are not neutral in this world, they are agents of influence and persuasion. Learners thus gain not just an abstract understanding of bias or media framing but a lived, experiential insight into how media narratives are constructed and weaponised (cf. Kolb's Experiential Learning Theory, 1984).

For example, a player who initially promotes press freedom may later choose to censor critical media outlets to protect national unity or ensure re-election. This introduces a crucial pedagogical moment: learners confront the ethical and systemic complexity of real-world media decision-making. The game allows space to explore why people in power might manipulate the media, and not just that they do. In doing so, it develops the kind of critical relational awareness that Hobbs (2010) identifies as a core goal of media pedagogy.

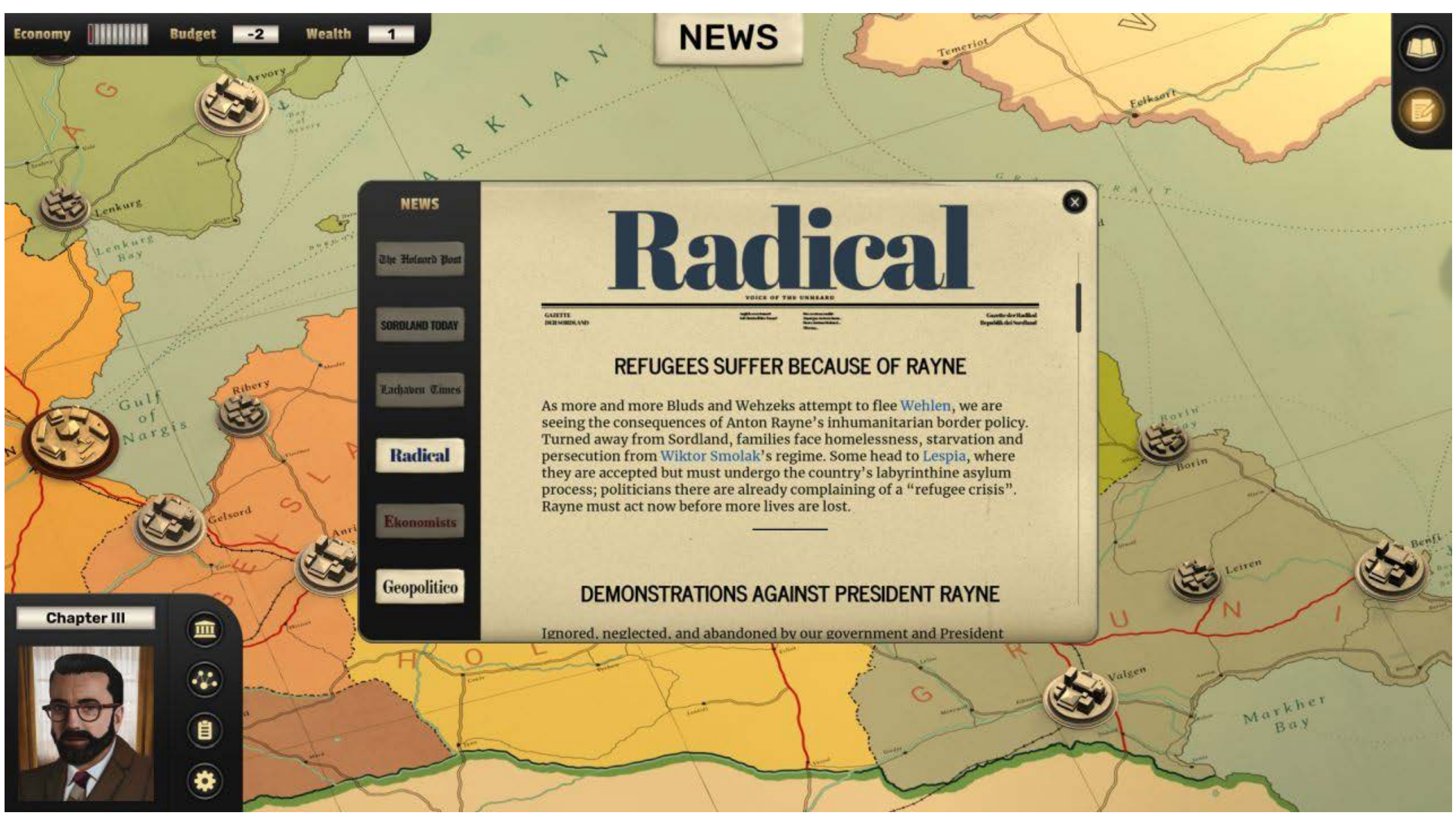

Figure 3: Screenshot from *Suzerain* (Torpor Games, 2020) illustrating multiple newspapers and the media's reactions to player choices.

### *5.1.2 Mapping Suzerain to the CLASS Framework*

The following breakdown analyses how Suzerain aligns with each of the five levers of the CLASS transdisciplinary pedagogical framework. This mapping highlights the game's pedagogical strengths and its capacity to support critical media literacy and systemic thinking in classroom contexts.

Table 1. Suzerain: CLASS levers mapped to application and media literacy and systems thinking integration

| **CLASS Lever** | **Application in Suzerain** | **Media Literacy and Systems Thinking Integration** |
|---|---|---|
| *Curious creativity* | Players explore a rich political fiction with multiple branching paths (authoritarianism, democratic reform, nationalism). Before each decision, the game evokes curiosity by prompting learners to consider the consequences of their options; after each outcome, it deepens that curiosity through the question: what if I had chosen differently? | Through sustained solo play and iterative replay ability, learners internalise how media narratives are constructed and contested, developing systemic curiosity by asking what different decisions would have changed across a complex political landscape. |
| *Lens & scope* | The game presents a wide ideological spectrum (communists, liberals, ultranationalists) alongside ethnic tensions, geopolitical complexity, and media bias, exposing learners to plural perspectives within a single simulated political system. | Prolonged immersion in a densely layered political world builds deep, personalised understanding of how media, power, and ideology interact as interdependent forces, rather than isolated variables. |
| *Agency* | Every player decision affects interconnected systemic variables (cabinet loyalty, press freedom, foreign alliances, public trust), placing learners in the role of a head of state navigating real ethical and political trade-offs. | Each unique sequence of choices gives the learner a personal causal history of how media and governance decisions ripple across a system, making systemic thinking felt rather than merely understood. |
| *Scaffolds* | In-game scaffolds including conversation trees, internal memos, and historical documents guide learners through Sordland's complex political history, with educators able to add further support through pre-reading, save files, and reflection prompts. | The layered narrative materials enable learners to build systemic and media literacy incrementally, moving from surface engagement with political events toward deeper understanding of institutional structures, causal relationships, and the role of information in shaping public discourse. |

| *Sandboxes* | The bounded fictional republic of Sordland allows learners to simulate different leadership styles and test ideological positions without real-world consequences, constrained by the game's constitution, budget, and political history. | The depth and duration of the experience creates the conditions for encountering delayed, unexpected systemic consequences, a form of learning that only emerges through extended individual engagement with a complex, consequential environment. |
|---|---|---|

### *5.1.3 Conclusion of case study 1*

*Suzerain* is a compelling example of how digital games can simulate the complexity of political systems and ethical decision-making. Its richly constructed fictional republic immerses players in scenarios that echo real-world dilemmas, including constitutional reform, economic policymaking, and international diplomacy. Distinct from strategy and simulation games that foreground statistical data and abstract visualisations (e.g. *Democracy* simulation series*,* Harris, 2005-2022), *Suzerain* privileges qualitative information and interpretive nuance. This design choice cultivates analytical reasoning, inferential skills, and the ability to navigate political ambiguity; essential competencies in both civic education and digital literacy.

Nevertheless, when considered through the CLASS lever lens & scope, *Suzerain* poses notable challenges for integration into formal curricula. A substantial cognitive investment is required to grasp the game's fictional lore, political history, and character relationships, time that might alternatively be directed towards the study of real-world political systems and events. A single playthrough requires approximately eleven hours, and meaningful exploration of divergent outcomes may demand upwards of 200 hours. Such demands exceed the practical time constraints of most higher education settings. Educators can address this by using the game's early hours as a shared case study, assigning segments as homework in a flipped-classroom model, or employing curated save files to focus on key decision points.

To maximise its pedagogical potential, educators must be equipped with structured guidance on how to implement *Suzerain* effectively in the classroom. Without such scaffolding, students may draw problematic conclusions, for instance, that populism or manipulation is a *necessary* path to success, if critical reflection is not explicitly facilitated.

Beyond its specific civic education context, the CLASS-based analysis of *Suzerain* offers a transferable methodology for educators in other disciplines. The approach taken here, mapping an existing IDN against the five levers to identify pedagogical strengths, limitations, and implementation strategies, can be applied to any narrative simulation, commercial or custom-built. An educator in environmental science could apply the same CLASS lens to an IDN simulating climate policy decisions; a business school instructor could use it to evaluate a corporate leadership simulation. The framework travels with the educator and context, not with the content.

### *5.2 Case Study 2*

The second case study examines the student-developed multiplayer narrative simulation Bezmarisk as a model of interactive media pedagogy. The experience is intentionally crafted for educational contexts while ensuring an engaging and enjoyable experience for participants. Through the lens of the CLASS Transdisciplinary Pedagogical Framework the analysis shows how the experience cultivates media literacy and systemic thinking.

#### *5.2.1 Introduction: Bezmarisk as speculative media pedagogy*

*Bezmarisk* is a live-action role-based simulation game situated in the fictional nation of Bezmarisk, a post-conflict state rebuilding its political infrastructure while facing escalating public security threats. Four learners assume the roles of newly appointed officials tasked with determining the ethical and practical use of surveillance technologies. The four must negotiate, argue, and due to timeboxing, ultimately decide which combinations of surveillance measures to adopt in response to shifting political, economic, and social pressures.

The game presents a carefully constructed environment where media technologies and political decision-making intersect. It draws directly on current global debates surrounding big data, algorithmic governance, predictive policing, and state surveillance. As such, *Bezmarisk* serves not just as a role-playing exercise but as a media literacy intervention, helping learners critically analyse the infrastructures and ideologies that underpin data-driven governance.

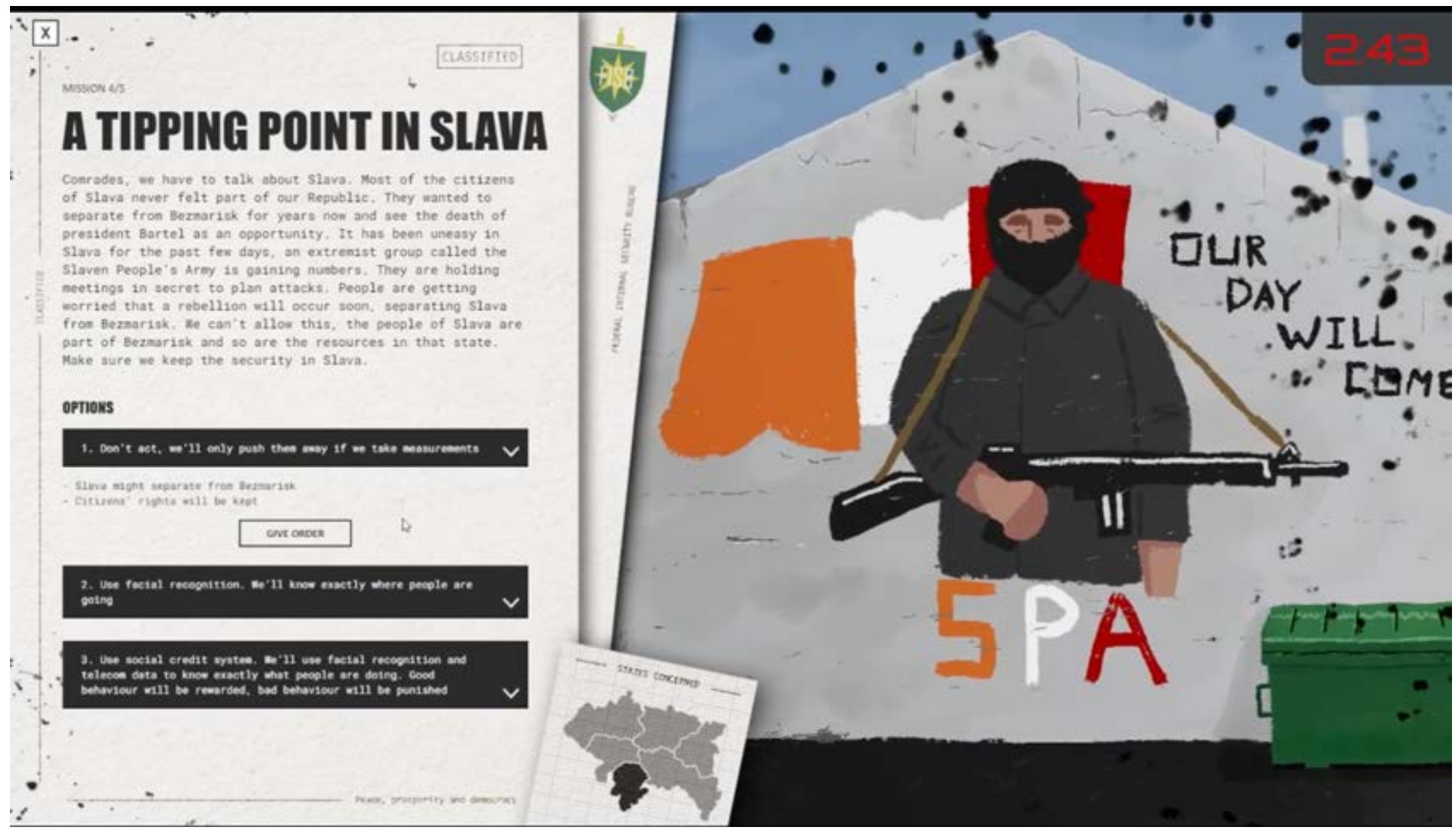

Figure 4: Screenshot from *Bezmarisk* showing the mission screen, which introduces a complex situation along with distinct response options.

*5.2.2 Media literacy and systems thinking*

Media literacy in *Bezmarisk* emerges not from decoding representations but from systemic engagement with media as tools of power. Unlike didactic instruction, the game structure demands systems thinking: players must consider how multiple variables interact (public opinion, crime rates, infrastructure, policy consequences) in an interdependent ecosystem. Furthermore, players assume roles with predefined positions and motivations, introducing perspective-taking, interpersonal friction, and elements of social deduction into the experience. The design enables a situated form of learning (Gee, 2004), in which understanding emerges from active participation in a complex, evolving context.

*5.2.3 Mapping Bezmarisk to the CLASS Framework*

Table 2. Bezmarisk: CLASS levers mapped to application and media literacy and systems thinking integration

| **CLASS Lever** | **Application in Bezmarisk** | **Media Literacy and Systems Thinking Integration** |
|---|---|---|
| *Curious creativity* | Players navigate a speculative surveillance scenario with hidden objectives and uncertain outcomes, triggering curiosity through the information gap created by each policy decision and its delayed reveal on the national map. | The information gap created by hidden role objectives and delayed map reveals generates collective curiosity, as players must anticipate how others' decisions will shape outcomes, fostering shared inquiry into media framing and systemic consequences. |
| *Lens & scope* | Four roles (human rights advocate, state security official, tech expert, and economist) each carry distinct objectives and worldviews, requiring learners to inhabit and negotiate across competing societal perspectives. | Negotiating across roles with genuinely conflicting objectives makes the systemic tension between media governance, civil liberties, and public safety experientially real, rather than abstractly understood. |
| *Agency* | Players collectively determine which surveillance measures their society adopts, with each policy combination leading to divergent national trajectories, reinforcing learners' sense of ethical authorship and systemic responsibility. | Collective decision-making under time pressure creates a shared sense of systemic authorship, as the group navigates ethical trade-offs between surveillance, security, and civil liberties with visible and immediate consequences for their society. |

| *Scaffolds* | Role briefings, a set of eight policy cards, and visual art-based feedback on societal change provide layered support, enabling learners at different levels to participate, interpret consequences, and build toward complex analysis. | The structured scaffolding enables learners at varying levels to engage meaningfully with complex governance dilemmas, with the facilitator guiding collective reflection on how shared decisions produced systemic consequences that no single role could have anticipated alone. |
|---|---|---|
| *Sandboxes* | The fictional nation of Bezmarisk provides a bounded, time-limited environment where learners can test surveillance strategies, experience unintended consequences, and reflect on systemic outcomes without real-world stakes. | The one-hour, multiplayer format creates a shared sandbox where teams observe systemic consequences together in real time, and the post-game debrief transforms individual role-based experience into collective media and systemic literacy. |

### *5.4 Conclusion of case study 2*

*Bezmarisk* illustrates the pedagogical strength of interactive storytelling in fostering critical media literacy and systems awareness. The current feature-complete prototype is a working example of a tailor-made systemic learning IDN that engages learners through roleplay and meaningful interaction. The IDN is suitable for different age groups and classroom contexts, and it is accessible and allows for online implementation. Mapped through the CLASS framework, the experience reveals how narrative simulations can be leveraged to teach ethical reasoning, collective decision-making, and the role of media infrastructures in shaping society. In a time of intensifying public discourse around data, surveillance, and democracy, such simulations offer timely and transformative educational interventions. Two experts in applied game design who were invited to review the experience as part of the student exam rated it as a highly valuable and well-executed prototype. While the game was successfully piloted with students from diverse backgrounds within a narrative design minor at a game-focused art university, further testing is required in secondary school contexts. Additionally, the development, implementation, and systematic evaluation of educator support materials will be essential for broader adoption. While promising, such approaches are not universally applicable across all educational contexts and require careful adaptation. Effective implementation depends on educator training and facilitation, particularly to support critical reflection and avoid misinterpretation of complex issues. Without appropriate scaffolding, there is a risk that learners may oversimplify or misread the systemic dynamics presented.

### *5.5 Comparison of the case studies*

Unlike longer-form narrative games such as *Suzerain*, which may require multiple hours of solo play, *Bezmarisk* is designed explicitly for a condensed one-hour session of multiplayer learning, making it highly adaptable to classroom time constraints. Its multiplayer, role-based structure fosters real-time interaction, allowing students to collaboratively explore ethical dilemmas while actively developing perspective-taking, negotiation, and communication skills through embodied role play.

While the constrained decision space of Bezmarisk, consisting of five scenario missions and limited reaction choices, may limit its sandbox appeal, this narrative framing also supports structured comparison and classroom discussion. In this sense, limited agency can function as a scaffold, enabling learners to understand better the systemic consequences of their choices within a simulated political ecosystem. The world of Bezmarisk is intentionally simplified: learners receive instruction sheets outlining their roles and (hidden) objectives, and a map visualises the consequences of decisions after each round. In contrast, Suzerain is a far more complex simulation, enabling a deeper exploration of political systems, media influence, and the tactical manoeuvring required to stay in charge. Instead of serving as alternatives, these experiences are most effective when used in tandem. Ideally, learners begin with the accessible framework of Bezmarisk, then progress to the more intricate opening of Suzerain , using their experiences to reflect on and compare both IDNs. This form of meta-reflection on how complexity is conveyed enhances media literacy by actively involving learners in evaluating the pedagogical approaches themselves, rather than simply subjecting them to predefined methods.

Together, these two case studies illustrate a spectrum that educators in any discipline can draw on when considering IDN-based learning. *Suzerain* represents one end: a richly complex, commercially available IDN that rewards deep engagement but requires significant scaffolding and time investment. *Bezmarisk* represents the other: a purpose-built, time-bounded experience designed for direct classroom use with minimal prior knowledge. Neither is universally better; the choice depends on the learning context, available time, and the educator's goals. What the CLASS framework provides is a common language for making that choice deliberately, and for designing the learning experience around it, regardless of the disciplinary setting. Educators working in environmental science, health, history, or business will face the same fundamental questions about curiosity, agency, scaffolding, scope, and safe experimentation that CLASS is designed to address. Together, these cases demonstrate that transferability lies not in replicating the narrative content, but in applying the CLASS design logic across contexts.

## 6. Conclusion: Stories and systems for the future of education

This paper has introduced a transdisciplinary pedagogical framework that integrates interactive digital narratives (IDNs), systems thinking, and design thinking to foster media literacy and systemic understanding in educational settings. By leveraging storytelling as a connective and cognitive scaffold, systemic learning IDNs create emotionally resonant, intellectually rigorous environments where learners can actively engage with the complexities of wicked problems.

CLASS makes a distinct contribution by bringing together interactive narrative design, systems thinking, and design thinking into a single structured framework explicitly oriented toward classroom use and transferable across disciplinary contexts.

The proposed CLASS transdisciplinary pedagogical framework supports the design and evaluation of such experiences, offering a structure that highlights the importance of curiosity, learner agency, perspective-taking, and systemic reflection. Rather than treating IDNs as standalone tools, we argue for their strategic integration into broader learning sequences that include reflection, mapping, discussion, and co-creation. When grounded in experiential learning theory (Kolb, 1984), such integration facilitates deeper internalisation of abstract systemic relationships through concrete narrative encounters.

Our case studies, *Suzerain* and *Bezmarisk*, demonstrate how IDNs can serve as participatory sandboxes, allowing learners to test choices, confront trade-offs, and uncover the ethical tensions inherent in complex systems such as media, governance, and civic life. These environments not only teach students to recognise interdependencies and power dynamics, but also help develop critical literacies, including civic empathy, strategic foresight, and reflective judgment.

As media landscapes grow more opaque and socio-political challenges more interconnected, education must move beyond content delivery toward systems-oriented sensemaking. Systemic learning IDNs offer a promising pathway. They are not just informational tools, but structured invitations to think differently; supporting learners in developing the critical, systemic, and ethical sensibilities needed to engage with complexity in thoughtful and responsible ways.

## 7. Future Work

Future research should examine the integration of systemic learning IDNs across diverse educational contexts, from higher education to early childhood settings. Designing age-appropriate IDNs that address complex socio-political themes can support critical thinking, ethical reflection, and media literacy from a young age. This includes helping learners recognise cognitive and algorithmic bias, interrogate oversimplified narratives, and assess the credibility of information. Expanding beyond digital formats, future work should also explore the potential of non-digital systemic narrative environments—such as board games, card systems, and analogue role-playing—as inclusive, screen-light alternatives, especially for younger learners. Emerging technologies like XR and AI offer further avenues for embodied, collaborative, and co-creative learning environments, while nature-based and evolutionary metaphors could inspire more accessible and politically neutral engagement with complex systems.

A robust framework for IDNs in education must balance three key modalities: learner-led creation, experiential consumption, and critical analysis. While this paper has focused on engaging with preexisting IDNs, the design and co-creation of narratives by learners fosters creativity, systems thinking, and deeper narrative engagement. Equally, the ability to critically analyse IDNs, including their structures, interaction mechanics, and embedded assumptions, is essential to developing sophisticated media literacy. Future work should also prioritise accessibility, ensuring inclusive design for neurodiverse learners, non-gamers, and individuals with varied abilities and literacy levels. Finally, ethical considerations and the potential unintended effects of transformative narratives, particularly on sensitive topics, must be addressed through iterative, co-designed, and evaluative processes that foreground cultural responsiveness and social responsibility.

## CRediT Authorship Contribution Statement

**Christian Roth:** Conceptualisation, Methodology, Project administration, Formal analysis, Investigation, Writing – original draft, Writing – review & editing. **Rahmin Bender-Salazar:** Conceptualisation, Methodology, Writing – original draft, Writing – review & editing. **Breanne Pitt:** Conceptualisation, Methodology, Writing – original draft, Writing – review & editing.

## Acknowledgements

The authors wish to thank the *Bezmarisk* student development team at HKU University of the Arts Utrecht for their prototype, which was used as a case study in this research. Furthermore, the authors benefited from participation in INDCOR COST Action CA18230 and CIRCUL'ARTs COST Action CA23117 for conceptualising this research and collaboration.

## Use of LLMs

LLMs (Grammarly and ChatGPT) were used in the preparation of the manuscript and were limited to improving grammar and flow and identifying opportunities to condense the text. All AI-assisted edits were subsequently reviewed, revised, and approved by all three authors. The analytical framing, theoretical argumentation, case study analysis, and conceptual contributions of the paper are entirely the authors' own.